\documentclass[11pt,journal,final]{IEEEtran}
\usepackage{graphicx}
\usepackage{subfigure}
\usepackage{amsmath}
\usepackage{amssymb}
\usepackage{amsthm}
\usepackage{color}
\usepackage{ifthen}
\newtheorem{thm}{Theorem}
\newtheorem{prop}{Proposition}
\newtheorem{lem}{Lemma}
\newtheorem{fact}{Fact}
\theoremstyle{definition}
\newtheorem{rem}{Remark}
\newtheorem{definition}{Definition}
\newtheorem{exam}{Example}

\title{Hamming Code for Multiple Sources} 
\author{ Rick Ma and Samuel~Cheng,~\IEEEmembership{Member,~IEEE} 
\thanks{R. Ma was with the Department of Mathematics at the Hong Kong University of Science and Technology, Hong Kong.}
\thanks{S. Cheng is with the School of Electrical and Computer Engineering, University of Oklahoma, Tulsa, OK, 74135 USA email: samuel.cheng@ou.edu.}
\thanks{A part of this work was presented in DCC 2010 and was submitted to ISIT 2010.}
\thanks{Manuscript received January 19, 2010} 
}

\begin{document}

\maketitle

\begin{abstract}
We consider Slepian-Wolf (SW) coding of multiple sources and extend the packing bound and the notion of perfect code from conventional channel coding to SW coding with more than two sources. We then introduce Hamming Codes for Multiple Sources (HCMSs) as a potential solution of perfect SW coding for arbitrary number of terminals. Moreover, we study the case with three sources in detail.  We present the necessary conditions of a perfect SW code and show that there exists infinite number of HCMSs. Moreover, we show that for a perfect SW code with sufficiently long code length, the compression rates of different sources can be trade-off flexibly. Finally, we relax the construction procedure of HCMS and call the resulting code generalized HCMS. We prove that every perfect SW code for Hamming sources is equivalent to a generalized HCMS.

\end{abstract}

\newcommand{\commenton}{\boolean{false}}

\section{Introduction}



SW coding refers to 
lossless distributed compression
of correlated sources. Consider $s$ correlated sources $X_1,X_2,\cdots,X_s$.
Assuming that encoding can only be performed separately that $s$ encoders can
see only one of the $s$ sources but the compressed sources are transmitted to a
base station and decompressed jointly. To the surprise to many researchers of
their time, Slepian and Wolf showed that  it is possible to have no loss in sum
rate under this constrained situation \cite{SlepianW:73}. That is, at least in
theory, it is possible to recover the source losslessly at the base station even
though the sum rate is barely above the joint entropy $H(X_1,X_2,\cdots,X_s)$. 

Wyner is the first who realized that by taking computed syndromes as the
compressed sources, channel 
codes can be used to implement SW coding \cite{Wyner:74}. The approach was rediscovered and popularized by Pradhan {\em et al.} more than two decades later \cite{PradhanR:99}.
Practical syndrome-based schemes for S-W coding using channel codes have then been studied in   \cite{StankovicLX:06,pradhan2005gcc,Ungerboeck:82,MarcellinF:90,YangCXZ:03,LiverisXG:03b,ChouPR:03,RebolloZ:03,FengZE:03,MitranB:02a,WangO:01,Servetto:02}. 
However, most work is restricted to the discussion of two sources  \cite{PradhanR:99,SchonbergRP:04,RimoldiU:97,Garcia-FriasZ:05,ChenHJL:07} except few exceptions \cite{Liveris2003,StankovicLX:06,ChengM:DCC2010}. 
In this paper,
we describe
a generalized syndrome based SW code and extend
the notions of a packing bound and a perfect code 
from regular channel coding to SW coding with arbitrary number of sources. 
Moreover,
we introduce {\em Hamming Code for Multiple Sources} (HCMSs) as a perfect code solution of SW coding for {\em Hamming sources} (c.f. Definition \ref{def:hamming_source}) and show that there exists infinite number of HCMSs for three sources. We then extend HCMS to a more inclusive form dubbed {\em generalized HCMS} and 
show the universality of generalized HCMS. Namely, any perfect SW code of Hamming sources can be reduced to a HCMS code through equivalent operation (c.f. Theorem \ref{thm:main}).  
 
%

The paper is organized as follows. In the next section, we will describe the problem setup and introduce definitions used in the rest of the paper. In Section \ref{sect:nullspace}, we will present a major lemma useful to the rest of paper. We will introduce HCMS in Section \ref{sect:HCMS}. In Section \ref{sect:threesources}, HCMS for three sources will be discussed in detail. Necessary conditions for perfect code will be given and the existence of HCMS for three sources will be shown. In Section \ref{sect:GHCMS}, we extend HCMS to generalized HCMS using the notion of {\em row basis matrix} as defined in Section \ref{sect:def}. 
In Section \ref{sect:main}, we will show the universality of generalized HCMS.

\section{Problem Setup}
\label{sect:def}
We will start with a general definition of syndrome based SW codes with multiple sources \cite{ChengM:DCC2010}. 

\begin{definition}[Syndrome based SW code]
\label{def:syndrome}
 A rate $(r_1,r_2,\cdots,r_s)$ {\em syndrome based SW code} for $s$ correlated
length-$n$ sources 
contains $s$ coding matrices $H_1, H_2, \cdots, H_s$ of sizes $m_1 \times
n, m_2 \times n, \cdots, m_s \times n$, where 
 $r_i = m_i / n$ for $i=1,2,\cdots,s$.
\begin{itemize}
 \item Encoding: The $i^{th}$ encoder compresses length-$n$ input ${\bf x}_i$
into ${\bf y}_i=H_i {\bf x}_i$ and transmit the compressed $m_i$ bits (with
compression rate $r_i=m_i/n$) to the base station
\item
Decoding: Upon receiving all ${\bf y}_i$, the base station decodes all sources
by outputting a most probable $\hat{\bf x}_1, \hat{\bf x}_2, \cdots, \hat{\bf
x}_s$ that satisfies $H_i \hat{\bf x}_i = {\bf y}_i, i=1,2,\cdots, s$. 
\end{itemize}
\end{definition}

For the ease of exposition, we will occasionally refer a compression scheme with its coding matrices $(H_1,\cdots,H_s)$ directly. Moreover, let us introduce the following definitions.
\begin{definition}[$(s,n,M)$-compression]
 We refer to {\em $(s,n,M)$-compression} as a SW code of $s$ length-$n$ source tuples with $M$ total compressed bits ($M=m_1+m_2+\cdots+m_s$).
\end{definition}
%
\begin{definition}[Compressible]
we will say the set of $s$-terminal source tuples
$S$ to be {\em compressible} by a SW code if any source tuple in $S$ can be reconstructed losslessly. Alternatively, we say the SW code can {\em compress} $S$. 
\end{definition}
Apparently, a SW code can compress $S$ 
if and only if its encoding map restricted to $S$ is injective (or 1-1). 

At one time instance, we call the correlation among different sources a {\em type-$0$ correlation} when all source bits from different terminal are the same. In general, we call the correlation a {\em type-$t$ correlation} if all source bits except $t$ of them are the same.   
For highly correlated source, we expect that the most probable sources are those with type-$0$ correlations for all $n$ time instances, and the next most probable sources are those with $n-1$ type-$0$ correlations and one type-$1$ correlation. We call these sources $s$-terminal {\em Hamming sources} of length $n$. Let us summarize the above in the following. 

\begin{definition}[Hamming sources]
\label{def:hamming_source}
A $s$-terminal {\em Hamming source} of length $n$ is $s$ length-$n$ source tuple that contains either 1) entirely type-$0$ correlations for $n$ time instances; or 2) type-$0$ correlations for $n-1$ time instances and one type-$1$ correlation.
\end{definition}

Let $S$ be
the set containing all $s$-terminal Hamming sources of length $n$.
By simple counting, the set $S$ 
has size $(s' n + 1) 2^n$, where $s'=s$ when the number of terminals $s > 2$ and 
$s'=1$ when $s=2$.
Thus, if $S$ is compressible by a SW code with $\mathcal{C}$ denoted as the set of all compressed outputs, we have a packing bound given by $ |\mathcal{C}| \ge (s' n+1) 2^n$. We call the code as {\em  perfect} if the equality in the packing bound is satisfied ($ i.e., |\mathcal{C}| = (s' n+1) 2^n$). The notion of perfectness can be generalized to any set $S$ of interests:
\begin{definition}[Perfect SW codes]
 A SW code is perfect if and only if $|\mathcal{C}| = |S|$.
\end{definition}
In other words, the encoding map restricted to $S$ is surjective (and injective also since $S$ is compressible) if the compression is perfect.
For the rest of the paper, let us restrict $S$ to denote 
the set containing all $s$-terminal Hamming sources of length $n$. Note that if 
$H_1,\cdots,H_s$ can compress $S$, the intersection of null spaces of $H_1,\cdots,H_s$ should only contain the all-zero vector. Otherwise, let $\bf x$ belong to the intersection and thus the $s$-tuple $(\overset{s}{\overbrace{{\bf x},\cdots,{\bf x}}}) \in S$ will have the same syndrome (all-zero syndrome) as  $(\overset{s}{\overbrace{{\bf 0},\cdots,{\bf 0}}}) \in S$. This contradicts with the assumption that $H_1,\cdots,H_s$ compresses $S$.

The following definitions are used to simplify the subsequent discussion. 

\begin{definition}[Hamming Matrix]
An $m$-bit Hamming matrix 
(of size $m \times (2^m-1)$)
consists of all 
nonzero column vectors of length $m$. 
Note that the parity check matrix of a Hamming code is a Hamming matrix.
\end{definition}


\begin{definition}[Surjective Matrix]
 A {\em surjective matrix} is a full rank fat or square matrix. 
\end{definition}


\ifthenelse{\commenton}
{I believe the following concepts is well-known already.
}
 
\begin{definition}[Row Basis Matrix]
Given a matrix $A$, 
we say a
surjective  
matrix $B$ is a {\em row basis matrix} of $A$,
 if
$row (B)= row (A)$,
where $row(A)$ denotes the row space of $A$, i.e., all linear combinations of rows of $A$.
\end{definition}

\begin{exam}[Row Basis Matrix] 
$\begin{pmatrix}
 100 \\
            011
 \end{pmatrix}
$
is a row basis matrix of both 
$ \begin{pmatrix}
       100 \\
100\\
111
      \end{pmatrix}$ and
$ \begin{pmatrix}
       100 \\
111\\
111
      \end{pmatrix}$.
\end{exam}
 
\begin{rem}[Row Basis Matrix ``Transform'']
\label{rem:total_compression}
There is a {\em unique} matrix $C$ s.t. $A=CB$
(because every row of $A$ can be decompose as a unique linear combination of $B$ since $B$ is full rank).
And there exists matrix $D$ s.t. $B=DA$ but $D$ is not necessary unique.
{\color{black}
Thus, given a vector $\bf v$, if we know $A{\bf v}$, we can compute $B{\bf v}$ (as $DA{\bf v}$). Similar,
we have $B{\bf v}$ given $A{\bf v}$.}
\end{rem}

\section{Null Space Shifting}
\label{sect:nullspace}

We will now introduce an important lemma that 
provides a powerful tool for our subsequent discussion.
In a nutshell, the lemma tells us that it is possible to tradeoff the compression rates of different source tuples by shifting a part of the null space of a coding matrix to another.

Our proof is based on analysis of the null spaces of the coding matrices. More precisely, we isolate the common space shared by all except one null spaces and decompose the null spaces as the direct sum ($\oplus$) of the common space and the residual space.


\begin{lem}
\label{lem:allocate}
Suppose $H_1,\cdots,H_s$ can compress $S$ and $\mbox{null} (H_i) = K \oplus N_i$ for all $i$ but a $r$ that $\mbox{null} (H_r) = N_r$, then 
matrices $H'_1,\cdots,H'_s$, with $\mbox{null} (H'_i) = K \oplus N_i$ for all $i$ but a $d$ that $\mbox{null} (H_d)=N_d$ can also compress 
$S$.

 
Furthermore, if all $H'_j$ are onto and $(H_1,\cdots,H_s)$ is a perfect compression, i.e.,
$(H_1,\cdots,H_s)$ restricted to $S$ is bijective, then $(H'_1,...,H'_s)$ is also a perfect compression. 
\end{lem}

Before the proof, we first notice that $N_r \cap K =\{{\bf 0}\}$. Otherwise $H_1,\cdots,H_s$ have common nonzero null vector that contradicts with $S$ being compressible by the code.
So the notation $K \oplus N_r$ is justified.
\begin{IEEEproof}
We have nothing to prove if $r=d$. For $r\neq d$, we can simply put $r=2$ and $d=1$ without losing generality.

 Define $S_+$ as $S+S=\{{\bf s}_1+{\bf s}_2| {\bf s}_1,{\bf s}_2\in S\}$. 
Note that
\begin{enumerate}
 \item $(H'_1,\cdots,H'_s)$ restricted to $S$ is 1-1 $\Leftrightarrow$ if $({\bf m}_1,\cdots,{\bf m}_s) \in S_+$ and $(H'_1 {\bf m}_1,\cdots,H'_s {\bf m}_s)=({\bf 0},\cdots,{\bf 0})$, then $({\bf m}_1,\cdots,{\bf m}_s)=({\bf 0},\cdots,{\bf 0})$.
\end{enumerate}
 So let $({\bf m}_1,\cdots,{\bf m}_s)\in S_+$ and $(H'_1{\bf m}_1,\cdots,H'_s {\bf m}_s)=({\bf 0},\cdots,{\bf 0})$.
By checking back the null spaces of $H'_j$,
we have ${\bf m}_1={\bf n}_1$,
             ${\bf m}_2={\bf k}_2+{\bf n}_2$,
             ${\bf m}_i={\bf k}_i+{\bf n}_i$,
where ${\bf n}_j\in N_j,$ and $\{{\bf k}_1,{\bf k}_2 \} \subset K$. So,
            ${\bf m}_1+{\bf k}_2={\bf n}_1+{\bf k}_2$,
             ${\bf m}_2+{\bf k}_2={\bf n}_2$, and 
              ${\bf m}_i+{\bf k}_2={\bf k}_2+{\bf k}_i+{\bf n}_i$.
By checking the null spaces of $H_i$, we find
$({\bf m}_1,\cdots,{\bf m}_s)+({\bf k}_2,\cdots,{\bf k}_2)\in (\mbox{null} H_1, \cdots, \mbox{null} H_s)$. 
As $({\bf k}_2,\cdots,{\bf k}_2)$ has all type-$0$ correlations, $({\bf m}_1,\cdots,{\bf m}_s) + ({\bf k}_2, \cdots, {\bf k}_2)$ is also in $S_+$.
        Since $(H_1,\cdots,H_s)$ restricted to $S$ is injective,
       by 1), we have   
                ${\bf m}_j={\bf k}_2$ for all $j$. In particular,
           ${\bf m}_1={\bf k}_2={\bf n}_1$, which implies ${\bf k}_2 \in N_1 \cap K = \{{\bf 0}\}$. That is, 
          ${\bf k}_2={\bf 0}$ and  thus ${\bf m}_j={\bf 0}$ for all $j$.
By 1) again, $(H'_1,\cdots,H'_s)$ restricted to S is injective.

For the second part, $(H_1,\cdots,H_s)$ restricted to $S$ is now surjective as well and hence
$(H_1,\cdots,H_s)$ per se is also surjective. Therefore all $H_j$ are full rank matrices. 
These imply all $H'_j$ have to be full rank as well. 
Furthermore, if all all $H'_j$ are full rank, 
the dimension of the target space of $(H'_1,\cdots, H'_s)= n s - \sum_{j=1}^s |\mbox{null}(H'_j)|= ns - \sum_{j=1}^s |\mbox{null}(H_j)|$, which equals to the the dimension of the target space of $(H1,\cdots,Hs)$. 
Since $(H_1,\cdots,H_s)$ restricted to $S$ is bijective, the dimension of the target space of  $(H_1,\cdots,H_s)$ (and hence that of $(H'_1,\cdots,H'_s)$) is $|S|$.
Since $(H'_1,\cdots,H'_s)$ restricted to $S$ has been proven to be injective already, it must be bijective as well.
\end{IEEEproof}


\section{Hamming Code for Multiple Sources}

\label{sect:HCMS}

Recall that $S$ is denoted as 
the set containing all $s$-terminal Hamming sources of length $n$. 
%
Let $M=m_1 + m_2 + \cdots + m_s$ be the total number of compressed bits. Then we have 
$|{\mathcal C}|= 2^M$ and thus when $s > 2$,
the equation for perfect compression becomes 
\begin{align}
2^n(sn+1)=2^M.                                
\label{eqn:1}
\end{align}
%
%

Since      $sn+1=2^{(M-n)}$,
$s$ obviously cannot be even. On the other hand, 
%
by Fermat's Little Theorem,
we have
           $1=2^{(s-1)} \pmod s$
for every odd prime $s>1$.
This
gives
an infinite number of solution to (\ref{eqn:1}). 




Now, we will present the main theorem that leads to HCMS.
 
\begin{thm}[Hamming Code for Multiple Sources]
\label{thm:HCMS}
For positive integers $s,n,M$ satisfy (\ref{eqn:1}) and $s > 2$, 
let $P$ be a Hamming matrix of size 
$(M-n) \times (2^{M-n}-1)=(M-n)\times(sn)$.

If $P$ can be partitioned 
into
\begin{align}
P=[Q_1,Q_2,\cdots,Q_s]                                 
\label{eqn:partition1}
\end{align}
\ifthenelse{\commenton}
{
(I don't use $P1$, $P2$... lest it should confuse with the column vectors of $P$)
}
such that  
each $Q_i$ is an $(M-n)\times n$ matrix and 
\begin{align}
Q_1+Q_2+\cdots+Q_s=0, 
\label{eqn:partition2}
\end{align}
and 
\begin{align}
 R=\begin{pmatrix}
    Q_1 \\
Q_2 \\
\cdots \\
Q_{s-1} \\
T
   \end{pmatrix}
\label{eqn:4}
\end{align} 
  is 
invertible for some arbitrary $T$, then
we have  
a set of $s$ parity check matrices
\begin{align}
\begin{pmatrix} G_1 \\ Q_1\end{pmatrix},
\begin{pmatrix} G_2 \\ Q_2\end{pmatrix},
\cdots,
\begin{pmatrix} G_s \\ Q_s\end{pmatrix}
\label{eqn:6}
\end{align}
that forms a perfect compression,            
where
$G_1, G_2,\cdots,G_s$ be any kind of row partition of $T$. That is, 
\begin{align}
 T=\begin{pmatrix}
G_1 \\
G_2 \\
\cdots \\
G_s  
 \end{pmatrix},
\label{eqn:5}
\end{align} 
%
and 
some $G_i$ can be chosen as a void matrix.
\end{thm}
\begin{IEEEproof}

For any $b$, $v_i\in Z_2^n$ s.t. $|v_1|+|v_2|+\cdots+|v_s|\le 1$, the input of correlated sources $[b+v_1, b+v_2, \cdots, b+v_s]$ will result in syndrome
 \begin{align*}
\left[
\begin{pmatrix} G_1 (b+v_1) \\ Q_1 (b+v_1)\end{pmatrix},
\begin{pmatrix} G_2 (b+v_2) \\ Q_2 (b+v_2)\end{pmatrix},
\cdots,
\begin{pmatrix} G_s (b+v_s)\\ Q_s (b+v_s)\end{pmatrix}\right]
 \end{align*} to be received at the decoder.
%
%
%
The decoder can then retrieve $(v_1,\cdots,v_s)$ since
\begin{align*}
& Q_1(b+v_1)+Q_2(b+v_2)+ \cdots+Q_s(b+v_s) \\
 = & Q_1(v_1)+\cdots+Q_s(v_s) & \mbox{(by (\ref{eqn:partition2}))} \\
   = & P\begin{pmatrix}
          v_1      \\
          v_2\\
\cdots       \\
          v_s 
      \end{pmatrix} &   \mbox{(by (\ref{eqn:partition1}))}
\end{align*}
and $P$ is bijective 
over the set of all length-$sn$ vectors with weight $1$. 
 
      After knowing $(v_1,\cdots,v_s)$, we can compute
   $G_1(b),\cdots,G_s(b)$ and $Q_1(b),\cdots,Q_s(b)$. Thus, we have
 \begin{align*}
\begin{pmatrix}
          Q_1    \\     
          Q_2      \\   
	\cdots\\
          Q_{s-1}\\  
          G_1     \\    
          G_2\\
          \cdots \\
          G_s
\end{pmatrix}(b)
=
\begin{pmatrix}
                   Q_1 \\
  Q_2 \\
\cdots \\
                Q_{s-1}    \\  T
\end{pmatrix} (b)
=R(b). 
 \end{align*}
\ifthenelse{\commenton}
{
(c.f. (\ref{eqn:5}), (\ref{eqn:4})).
}
As $R$ is invertible, we can recover $b$ and hence the correlated sources
$[b+v_1, b+v_2, \cdots, b+v_s]$.
 
\end{IEEEproof}


\begin{rem}[SW coding of three sources of length-$1$]
\label{rem:counter}
Apparently, HCMS only exists if the $(s-1)(M-n) \le n$, otherwise the required height of $T$ will be negative. For example,
let $s=3$, $n=1$, $M=3$. Even though the parameters satisfy \eqref{eqn:1},
we will not have HCMS because  $n-(s-1)(M-n)=-3$. 
However, a perfect (trivial) SW code actually exists in this case, the parity check matrices for all three terminals are simply the scalar matrix $\begin{pmatrix}
1 \end{pmatrix}$. 

\end{rem}

From Remark \ref{rem:counter}, we see that HCMS cannot model all perfect codes that can compress $s$-terminal Hamming sources. It turns out that we can modify HCMS slightly and the extension will cover all perfect SW codes for Hamming sources. 
We will delay this discussion to Section \ref{sect:main}. In the next section, we will first discuss HCMS for three sources in detail. 

\section{HCMS for Three Sources}
\label{sect:threesources}
\subsection{Necessary Conditions}
Now, let us confine to the case with only three encoders, i.e., $s=3$. Let $(H_1,H_2,H_3)$ be a
perfect compression for $\mathcal{S}$. We have 
\begin{equation}
\mbox{null} H_1 \cap \mbox{null} H_2 \cap \mbox{null} H_3=\{ {\bf 0} \}.
\label{eqn:star} 
\end{equation}
 Hence we can
decompose the null spaces into $U\oplus K_2\oplus K_3, V\oplus K_1\oplus K_3,
W\oplus K_1\oplus K_2$, where $K_i=\mbox{null} H_j\cap \mbox{null} H_k, (i,j,k)\in\{(1,2,3), (2,3,1), (3,2,1)\}$. Notice that $(U\oplus K_2\oplus K_3) \cap K_1=\{{\bf 0}\}$ by (\ref{eqn:star}). We also have
$(U\oplus K_2\oplus K_3\oplus K_1)\cap W=\{{\bf 0}\}$ (because
$\forall {\bf u}\in U$ ,$\forall {\bf k}_1 \in K_1$ , $\forall {\bf k}_2 \in K_2 $, $\forall {\bf k}_3 \in K_3$, and $\forall {\bf w} \in W$,  
 ${\bf u}+{\bf k}_1+{\bf k}_2+{\bf k}_3={\bf w} \Rightarrow
{\bf u}+{\bf k}_2+{\bf k}_3={\bf k}_1+{\bf w}\in \mbox{null} H_1\cap \mbox{null} H_3=K_2 \Rightarrow {\bf k}_1={\bf 0}$ and ${\bf w}={\bf 0}$). By the symmetry among
$U,V,W$, we also have $(C\oplus A)\cap B =\{{\bf 0}\}$ for any $A,B\in \{U,V,W\}$ and $A \neq B$, where
$C= K_1\oplus K_2\oplus K_3$. By Lemma 1, the perfect compression is equivalent to full rank matrices with null space $C\oplus U, C\oplus V$, and $W$. 


Let us denote the dimensions of $U,V,W$, and $C$, as $u,v,w$, and $c$, respectively. Then we have the following lemma.
\begin{lem}
\label{lem:range}
 With $u,v,w$, and $c$ described above, if the code can compress $S$ and  is perfect,
then $3 n - 2 M \le c \le 3n-2M +3$, where $M=m_1+m_2+m_3$. Moreover,
$M -n -2  \le u,v, w \le M-n$.
\end{lem}
\begin{IEEEproof}
 Even if the rest of the sources are known exactly, the correlation specified by $S$ implies that there can be $n+1$ possibilities for the remaining source. Therefore, $2^{m_i} \ge n+1$ for $i=1,2,3$. Denote $D_i$ as the dimension of the null space of $H_i$. Then, 
$
D_i = n - m_i \le n-\log_2(n+1)<  n - \log_2(n+ 1/3) = n-\log_2(3n+1) + \log_2 3.
$
From the packing bound, if the code is perfect, we have $2^M = (3n+1) 2^n$. Therefore, $D \le n - (M-n) + \log_2 3$ and thus $D \le 2n - M +1$, where the second inequality holds because $D$, $n$, and $M$ are all integers.

Given the three null spaces to be $ C \oplus U$, $C \oplus V$, and $W$, then we have $ c + v \le 2 n - M  +1 $, $c + u \le 2n - M +1$, and $c + w\le 2n - M +1$. Note that the last inequality holds 
since Lemma \ref{lem:allocate} tells us that full rank matrices with null spaces $C \oplus U$, $V$, $C \oplus W$ also compress $S$ perfectly.
Moreover, since 
$(C \oplus A) \cap B = \left\{  {\bf 0} \right\}$ 
 for any $A,B \in \{U,V,W\}$ and $A \neq B$, 
we have $c+ u + v\le n$, $c+u+w \le n$, and $c+v+w \le n$. And the  total dimensions of the null spaces $= 3 n - M = c + u + c+v +w$. Thus $3n - M - (c+u+v) = c+w $ and $2n-M \le c+w$. Similarly, we have $2n - M \le c+u$ and $2n-M \le c+v$. In summary, we have 
\begin{equation}
2 n - M \le c + a \le 2n-M+1,
\label{eqn:ca}
\end{equation} $a \in \{ u,v,w \}$. 
Since $w=3 n - M -(c+u) - (c+v)$, by 
(\ref{eqn:ca}), $ 3n -M -2 (2n-M +1) \le w \le 3n -M -2(2n-M)$ and thus $M-n -2 \le w \le M-n$.  Similarly, we have $M-n-2 \le u,v \le M-n$.

Now, substituting $M-n -2 \le w \le M-n$ back to (\ref{eqn:ca}), we have $3n -2 M \le c \le 3n - 2M +3$ as desired.
\end{IEEEproof}

To summarize from Lemma \ref{lem:range}, given $n$ and $M$, there are only four cases for different values of $c,u,v$, and $w$ assuming $u \ge v \ge w$ as shown in Table \ref{table:4cases}. 

\begin{table}
\caption{
Feasible values of $c,u,v$, and $w$}
{\small
\begin{center}
\begin{tabular}{c | c| c| c}
$c$ & $u$ & $v$ & $w$ \\ \hline 
$3n-2M+3$ & $M-n-2$ & $M-n-2$ & $M-n-2$ \\
$3n-2M+2$ & $M-n-1$ & $M-n-1$ & $M-n-2$ \\
$3n-2M+1$ & $M-n$ & $M-n-1$ & $M-n-1$ \\
$3n-2M$ & $M-n$ & $M-n$ & $ M-n$
 \end{tabular}
\end{center}}
\label{table:4cases}
\end{table}

Moreover, the perfectness condition,  
$2^M = (3n+1) 2^n$, 
turns out to be restrictive enough to confine $M$ and $n$ into some limited possibilities as to be describe in the following Lemma.

\begin{lem}
All positive integers $n$ and $M$ that satisfy $2^M = 2^n (3n+1)$ 
have the forms $(2^{2a}-1) /3$ and $2a+(2^{2a}-1)/3$, respectively, for some positive integer $a$. 
\label{lem:Mn}
\end{lem}
\begin{IEEEproof}
 It is easy to verify that $2^{2 a} \equiv 1 \pmod 3$ and $2^{2 a + 1} \equiv 2 \pmod 3$ for any positive integer $a$. Moreover, since both $M$ and $n$ are positive integers, $3 n+ 1 = 2^{M-n}$ is a positive integer as well. This implies $2^{M-n} \equiv 1 \pmod 3$ and thus $M-n$ has to be even. Let $M-n = 2 a$ for some positive integer $a$, then we can rewrite $n$ and $M$ as $a$: $n(a) = (2^{2a} -1)/3$ and $M(a) = 2 a + n(a)$.
\end{IEEEproof}

It is interesting to point out that $M$ has to be divisible by $3$. It can be proved using simple induction and we will skip the proof here.
\begin{lem}
 $M(a) = (2^{2a}-1)/3+2a$, $a \in \mathbb{Z}^+$, defined in Lemma \ref{lem:Mn} is divisible by $3$.
\end{lem}

We list in Table \ref{table:nM} the first five possible values of $M$ and $n$. $M-n$ and $3n-2M$ are also included for convenience. 

\begin{table}
\caption{Values of $n$,$M$, $M-n$, $3n-2M$ for $a=1,2,3,4,5$.}
{\small
\begin{center}
\begin{tabular}{c | c | c | c| c}
$a$ & $n$ & $M$ & $M-n$ & $3n - 2M$ \\ \hline
$1$ & $1$ & $3$ & $2$ & $-3$ \\ 
$2$ & $5$ & $9$ & $4$ & $-3$ \\
$3$ & $21$ & $27$ & $6$ & $9$ \\
$4$ & $85$ & $93$ & $8$ & $69$ \\
$5$ & $341$ & $351$ & $10$ & $321$ 
\end{tabular}
\end{center}}
\label{table:nM}
\end{table}

Note that for both $n=1$ and $n=5$, $3n-2M = -3$. Thus, only the first case described in Table \ref{table:4cases} will be possible (i.e. $c=0$). 

 By Lemma \ref{lem:allocate}, the null space contributed by $C$ can be reallocated to different terminals arbitrarily and yet the resulting code will still compress $S$ and be perfect. 

For example, for $n=21$ and $M=27$, possible values of $c,u,v,w$ are
$12,4,4$, and $4$, respectively. This results in an asymmetric code with $m_1=21-12-4=5$, $m_2=5$, and $m_3=17$. If this code compresses $S$, we can reallocate $4$ dimensions of null spaces each from $H_1$ and $H_2$ to $H_3$ and result in a symmetric code that can compress $S$ as well. 

From the above discussion, we see that $c \sim 3 n - 2 M$ increases exponentially with $a$ wherea  $u,v,w \sim M-n$ only increases linearly. Therefore, for sufficiently large $n$, $c$ will always be large enough that we can rearrange any perfect code to another asymmetric perfect code of desired 
rates 
by allocating $C$ among different coding matrices. 

\subsection{Existence of HCMS}

We have not yet shown that any HCMS exists. Actually, the authors are not aware of any prior work that reported perfect SW codes with more than two sources in the literature. 
From Remark \ref{rem:counter}, we see that HCMS cannot model the trivial case for $a=1$ ($n=1$ and $M=3$) even though by definition the trivial code (coding matrices equal to scalar identity for all three sources) is perfect. For $a=2$ ($n=5$ and $M=9$), HCMS also does not exist since $n - (s-1) (M-n) = 5-2(4) = -3 < 0$ (c.f. Remark \ref{rem:counter}). However, there is actually no perfect code exists at all in this case as concluded in the following proposition. 

\begin{prop} [No perfect code for $a=2$]
\label{prop:a=2}
There does not exist perfect code for SW coding of three length-$5$ sources. 
\end{prop}
\begin{IEEEproof}
 See Appendix.
\end{IEEEproof}

Even though HCMS does not exist for $a=1$ and $a=2$, it is possible to show that HCMS exists for $a \ge 3$.
In the following, we will first show that HCMS for three sources exists for $a\ge 3$ using Theorem \ref{thm:HCMS}. 
\begin{prop} [HCMS exists for $a=3$]
\label{prop:a=3}
 Now we will show that perfect compression exists for $s=3$ and $a= 3$.
For $a=3, n=21$ and $M=27$, we let
 
\small
$$
Q_1 =
\begin{pmatrix}
1 \; 0 \; 0 \; 0 \; 0 \; 0 \; 1 \; 0 \; 0 \; 0 \; 0 \; 1 \; 1 \; 1 \; 0 \; 1 \; 1 \; 0 \; 0 \; 0 \; 0 \\
0 \; 1 \; 0 \; 0 \; 0 \; 0 \; 1 \; 1 \; 0 \; 0 \; 0 \; 0 \; 1 \; 0 \; 0 \; 0 \; 0 \; 0 \; 1 \; 1 \; 1 \\
0 \; 0 \; 1 \; 0 \; 0 \; 0 \; 0 \; 1 \; 1 \; 0 \; 0 \; 0 \; 0 \; 1 \; 1 \; 1 \; 0 \; 1 \; 0 \; 1 \; 1 \\
0 \; 0 \; 0 \; 1 \; 0 \; 0 \; 0 \; 0 \; 1 \; 1 \; 0 \; 0 \; 0 \; 1 \; 0 \; 0 \; 1 \; 1 \; 1 \; 1 \; 0 \\
0 \; 0 \; 0 \; 0 \; 1 \; 0 \; 0 \; 0 \; 0 \; 1 \; 1 \; 0 \; 1 \; 0 \; 1 \; 1 \; 0 \; 1 \; 1 \; 1 \; 1 \\
0 \; 0 \; 0 \; 0 \; 0 \; 1 \; 0 \; 0 \; 0 \; 0 \; 1 \; 1 \; 0 \; 0 \; 1 \; 0 \; 0 \; 1 \; 1 \; 0 \; 1
\end{pmatrix},
$$

$$
Q2=     
\begin{pmatrix}
0 \; 0 \; 0 \; 1 \; 0 \; 1 \; 1 \; 0 \; 1 \; 1 \; 1 \; 1 \; 0 \; 1 \; 0 \; 0 \; 0 \; 1 \; 1 \; 1 \; 1 \\
1 \; 0 \; 0 \; 0 \; 1 \; 0 \; 1 \; 1 \; 0 \; 1 \; 1 \; 1 \; 1 \; 0 \; 1 \; 1 \; 1 \; 1 \; 0 \; 0 \; 0 \\
0 \; 1 \; 0 \; 0 \; 0 \; 1 \; 1 \; 1 \; 1 \; 0 \; 1 \; 1 \; 1 \; 0 \; 0 \; 0 \; 0 \; 0 \; 1 \; 0 \; 1 \\
1 \; 0 \; 1 \; 0 \; 0 \; 0 \; 1 \; 1 \; 1 \; 1 \; 0 \; 1 \; 0 \; 1 \; 1 \; 1 \; 0 \; 0 \; 1 \; 1 \; 1 \\
0 \; 1 \; 0 \; 1 \; 0 \; 0 \; 1 \; 1 \; 1 \; 1 \; 1 \; 0 \; 0 \; 0 \; 1 \; 0 \; 1 \; 1 \; 0 \; 1 \; 1 \\
0 \; 0 \; 1 \; 0 \; 1 \; 0 \; 0 \; 1 \; 1 \; 1 \; 1 \; 1 \; 1 \; 1 \; 0 \; 1 \; 0 \; 1 \; 1 \; 1 \; 0
\end{pmatrix},
$$
\normalsize
and
\small
$$ 
 Q_3=    
\begin{pmatrix}
1 \; 0 \; 0 \; 1 \; 0 \; 1 \; 0 \; 0 \; 1 \; 1 \; 1 \; 0 \; 1 \; 0 \; 0 \; 1 \; 1 \; 1 \; 1 \; 1 \; 1  \\
1 \; 1 \; 0 \; 0 \; 1 \; 0 \; 0 \; 0 \; 0 \; 1 \; 1 \; 1 \; 0 \; 0 \; 1 \; 1 \; 1 \; 1 \; 1 \; 1 \; 1  \\
0 \; 1 \; 1 \; 0 \; 0 \; 1 \; 1 \; 0 \; 0 \; 0 \; 1 \; 1 \; 1 \; 1 \; 1 \; 1 \; 0 \; 1 \; 1 \; 1 \; 0  \\
1 \; 0 \; 1 \; 1 \; 0 \; 0 \; 1 \; 1 \; 0 \; 0 \; 0 \; 1 \; 0 \; 0 \; 1 \; 1 \; 1 \; 1 \; 0 \; 0 \; 1  \\
0 \; 1 \; 0 \; 1 \; 1 \; 0 \; 1 \; 1 \; 1 \; 0 \; 0 \; 0 \; 1 \; 0 \; 0 \; 1 \; 1 \; 0 \; 1 \; 0 \; 0  \\
0 \; 0 \; 1 \; 0 \; 1 \; 1 \; 0 \; 1 \; 1 \; 1 \; 0 \; 0 \; 1 \; 1 \; 1 \; 1 \; 0 \; 0 \; 0 \; 1 \; 1
\end{pmatrix}.
$$ 
\normalsize

It is a little bit laborious but straightforward to see that
$P=[Q_1 Q_2 Q_3]$ is a 6-bit Hamming matrix 
and $Q_1+Q_2+Q_3=0$. So we have \eqref{eqn:partition1} and \eqref{eqn:partition2} already. For \eqref{eqn:4}, let
$U$ and $V$ be the matrices obtained by truncating the last $9$ columns of $Q_1$ and $Q_2$:
 $$
U=   
\begin{pmatrix}
1 \; 0 \; 0 \; 0 \; 0 \; 0 \; 1 \; 0 \; 0 \; 0 \; 0 \; 1  \\
0 \; 1 \; 0 \; 0 \; 0 \; 0 \; 1 \; 1 \; 0 \; 0 \; 0 \; 0  \\
0 \; 0 \; 1 \; 0 \; 0 \; 0 \; 0 \; 1 \; 1 \; 0 \; 0 \; 0   \\
0 \; 0 \; 0 \; 1 \; 0 \; 0 \; 0 \; 0 \; 1 \; 1 \; 0 \; 0   \\
0 \; 0 \; 0 \; 0 \; 1 \; 0 \; 0 \; 0 \; 0 \; 1 \; 1 \; 0 \\
0 \; 0 \; 0 \; 0 \; 0 \; 1 \; 0 \; 0 \; 0 \; 0 \; 1 \; 1   \\
\end{pmatrix}
 $$
 and
 $$
V = 
\begin{pmatrix}
0 \; 0 \; 0 \; 1 \; 0 \; 1 \; 1 \; 0 \; 1 \; 1 \; 1 \; 1 \\ 
1 \; 0 \; 0 \; 0 \; 1 \; 0 \; 1 \; 1 \; 0 \; 1 \; 1 \; 1 \\
0 \; 1 \; 0 \; 0 \; 0 \; 1 \; 1 \; 1 \; 1 \; 0 \; 1 \; 1  \\
1 \; 0 \; 1 \; 0 \; 0 \; 0 \; 1 \; 1 \; 1 \; 1 \; 0 \; 1   \\
0 \; 1 \; 0 \; 1 \; 0 \; 0 \; 1 \; 1 \; 1 \; 1 \; 1 \; 0  \\
0 \; 0 \; 1 \; 0 \; 1 \; 0 \; 0 \; 1 \; 1 \; 1 \; 1 \; 1 \\
\end{pmatrix}.
 $$
 
We have $V=[0| I_6]+ KU$, where
 $$
K= 
\begin{pmatrix}
0 \; 0 \; 0 \; 1 \; 0 \; 1 \\
1 \; 0 \; 0 \; 0 \; 1 \; 0 \\
0 \; 1 \; 0 \; 0 \; 0 \; 1 \\
1 \; 0 \; 1 \; 0 \; 0 \; 0 \\
0 \; 1 \; 0 \; 1 \; 0 \; 0 \\
0 \; 0 \; 1 \; 0 \; 1 \; 0
\end{pmatrix}.
 $$
 Hence 
$\begin{pmatrix}
 Q_1 \\Q_2 
 \end{pmatrix}$ is a  full rank matrix with pivots on the first $12$ columns.
 Therefore,
$R=\begin{pmatrix}
Q_1 \\ Q_2 \\ T
\end{pmatrix}$ is a $21 \times 21$ invertible matrix where
                $T=[0| I_9]$.
 By Theorem \ref{thm:HCMS},
coding matrices
$
\begin{pmatrix}
               G_1  \\
               Q_1 
\end{pmatrix},
\begin{pmatrix}
               G_2  \\
               Q_2 
\end{pmatrix}
,\begin{pmatrix}
               G_3  \\
               Q_3 
\end{pmatrix}
$ form a perfect compression with $G_i$ defined in \eqref{eqn:5}. 
\end{prop}

\begin{thm} [HCMS exists for all $a \ge 3$]
 For $s=3$, there is a perfect compression for $n$ and $M$ that satisfy
\begin{align}
3n(a)+1=2^{2a},                                        \label{eqn:7}    \\     
M(a)-n(a)=2a, \label{eqn:8}                                                   
\end{align}
 where $a$ is any integer larger than or equal to $3$.
%
\end{thm}


\begin{IEEEproof}
From Proposition \ref{prop:a=3}, we have shown that there exists HCMS for three sources when $a=3$.
Now we will show by induction that there is perfect compression for all $a\ge 3$.
Suppose we have a partition of 
a Hamming matrix $P$ of size $2a \times (2^{2a}-1)$ (formed by all non-zero length-$2a$ column vectors) as
 $P=[A B C]$ such that $A+B+C=0$,    
 and
$\begin{pmatrix}
A\\B  
 \end{pmatrix}
$ forms a full rank matrix with pivots on the first $4a$ columns whenever $3 \le a \le k$.
By Theorem \ref{thm:HCMS}, perfect compression can be built by choosing $T=[0, I_{n(a)-4a}]$. 
 
We will show that the statement is also true for $k+1$.

{\color{black}

Let $P$ be the $2k$-bit Hamming matrix and $A,B,C$ be its partition with
properties described above.
Let
$ {\bf u}= \begin{pmatrix}
      1 \\0 
     \end{pmatrix}, 
{\bf v}= 
\begin{pmatrix}
 0 \\1
\end{pmatrix},
{\bf w}=
\begin{pmatrix}
 1 \\ 1
\end{pmatrix}.
$
Let $A_i$, $B_i$, $C_i$ be the $i$-th column of $A,B,C$ respectively. We define
\begin{align*}
A_+ = 
\left(
\begin{matrix}
{\bf 0}  \\
        {\bf u} 
\end{matrix} \; \cdots \;
\begin{matrix}
A_j &   A_j   & A_j &   A_j  \\
 {\bf 0} &    {\bf u} &    {\bf v} &     {\bf w} 
\end{matrix} \; \cdots \right) \\
B_+ = 
\left(
\begin{matrix}
{\bf 0}   \\
         {\bf v} 
\end{matrix} \; \cdots \;
\begin{matrix}
  B_j   &  B_j &    B_j  &  B_j  \\
   {\bf 0}   &  {\bf v}  &  {\bf w}   &  {\bf u}  
\end{matrix} \; \cdots \right) \\
C_+ = 
\left(
\begin{matrix}
{\bf 0}   \\
        {\bf w}  
\end{matrix} \; \cdots \;
\begin{matrix}
 C_j &  C_j &  C_j &  C_j  \\
  {\bf 0} &   {\bf w} &   {\bf u} &   {\bf v}  
\end{matrix} \; \cdots
\right)
 \end{align*}
where $j$ runs from $1$ to $n=(2^{2k}-1)/3$.
 
It is easy to verify that
$P_+ = [A_+ B_+ C_+]$ is a $2(k+1)$-bit Hamming matrix consisting of 
$3+(4(2^{2k}-1))=2^{2(k+1)}-1$ different non-zero column vectors of length $2(k+1)$. 
And obviously
$A_+ + B_+ + C_+=0$ 
 as
$A+B+C=0$ and ${\bf u}+{\bf v}+{\bf w}={\bf 0}$.
 
Lastly we permutated the columns of $A_+$, $B_+$ and $C_+$ simutanteously (keeping their sum  zero) such that the first $4(k+1)$ columns are 
\begin{align}
A_+ &= 
\begin{pmatrix}
A_1 & A_2 & \cdots & A_{2k} &  A_1 &  A_2 & A_3 &  A_4 & \cdots   \\
         {\bf 0} &   {\bf 0}&  \cdots &  {\bf 0} &       {\bf v} &    {\bf w} &   {\bf w} &   {\bf u} & \cdots
\end{pmatrix} \nonumber \\
B_+ &= 
\begin{pmatrix}
B_1 & B_2 & \cdots &B_{2k} & B_1 & B_2 & B_3 & B_4 & \cdots \\
         {\bf 0}&    {\bf 0}& \cdots  &  {\bf 0}&       {\bf w}  &  {\bf u} &   {\bf v} &    {\bf w}& \cdots
 \end{pmatrix} \nonumber \\
C_+ &= 
\begin{pmatrix}
 C_1 & C_2& \cdots&  C_{2k}&  C_1&  C_2&  C_3 &  C_4& \cdots \\
         {\bf 0}   & {\bf 0} & \cdots  & {\bf 0}      &  {\bf u}  &  {\bf v}   &  {\bf u}    & {\bf v}& \cdots                   
\end{pmatrix} 
 \end{align}

 
Notice that
$
\begin{pmatrix}
 A_1 \\B_1
\end{pmatrix},
\begin{pmatrix}
 A_2 \\B_2
\end{pmatrix},
\cdots,
\begin{pmatrix}
 A_{2k} \\B_{2k}
\end{pmatrix}
$ 
are linear independent by induction assumption and
 \begin{align*}
\begin{pmatrix}
{\bf v} &  {\bf w} &  {\bf w} &  {\bf u} \\
{\bf w} & {\bf u} & {\bf v} & {\bf w}   
\end{pmatrix}
=
\begin{pmatrix}
   0 & 1 & 1 &  1 \\
    1 & 1 & 1 & 0 \\
    1 & 1 & 0 & 1  \\
     1 & 0 & 1  & 1  
\end{pmatrix}
 \end{align*}
are also linear independent. Therefore the first $4(k+1)$ column vectors
of $\left[\begin{smallmatrix} A_+ \\ B_+ \end{smallmatrix}\right]$ are linearly independent. Hence $\left[ \begin{smallmatrix} A_+ \\ B_+ \end{smallmatrix} \right]$ is a full rank matrix with pivots
on the first $4(k+1)$ columns. By Theorem \ref{thm:HCMS}, perfect compression can be built
by choosing $T=[0, I_{n(k+1)-4(k+1)}]$.
 
 
}
\end{IEEEproof}


 

\ifthenelse{\commenton}
{
  After this email, I will send you the existence of perfect compression for $s=3$ (for all $a\ge 3$). Then I will send the $s=5$
case. Lastly I will explain the uniqueness of HCMS++.
}
 
\section{Generalized HCMS}
 \label{sect:GHCMS}

Now, we will extend HCMS so that it will cover the trivial case described in Remark \ref{rem:counter}.
The main idea of generalized HCMS is to ``loosen''
the condition in (\ref{eqn:4}) using the notion of row basis matrices defined in Section \ref{sect:def}. 
Let
$Y$, a $d \times n$ matrix, be 
a row basis matrix of    
$\begin{pmatrix}
 Q_1 \\
Q_2 \\
\cdots \\
Q_{s-1} \\
\end{pmatrix}$, where $Q_1,\cdots,Q_s$ is a partition of a Hamming matrix satisfying \eqref{eqn:partition1} and \eqref{eqn:partition2}.
%
Since $Y$ is a surjective 
matrix, there apparently exists $T$
s.t.
$     
R= 
\begin{pmatrix}
 Y \\
 T
\end{pmatrix}
$
is an $n \times n$ invertible matrix. 

\begin{thm}[Generalized HCMS]

Let $G_i, i=1,\cdots,s$ be any row partition of $T$ as in (\ref{eqn:5}) 
and $C_i$ be a row basis matrix of $Q_i$ for $i=1,\cdots,s$.
 Then 
a set of parity matrices 
$\begin{pmatrix}
 G_1 \\
 C_1
 \end{pmatrix}, 
\begin{pmatrix}
  G_2  \\
 C_2  
 \end{pmatrix}, 
\cdots,
\begin{pmatrix}
 G_s \\
  C_s
 \end{pmatrix} 
$
forms a compression for the set of $s$-terminal Hamming sources of length $n$.
Moreover,
%
the compression will be perfect if
$d_1+d_2+...+d_s+(n-d)=M$,
where $d_i$ is the number of row of $C_i$.
\end{thm}
 
\begin{IEEEproof}
{\color{black}
Let $|\cdot|$ be the function that maps an element in ${\mathbb Z}_2^n$ to its norm in $\mathbb Z$ by counting the number of nonzero components, e.g., $|(1,1,0,1)|=3$.
For any ${\bf b}$, ${\bf v}_i\in Z_2^n$ s.t. $|{\bf v}_1|+|{\bf v}_2|+\cdots+|{\bf v}_s|\le 1$, the input of correlated sources $[{\bf b}+{\bf v}_1, {\bf b}+{\bf v}_2, \cdots, {\bf b}+{\bf v}_s]$ will result in syndrome
 \begin{align*}
\left[
\begin{pmatrix} G_1 ({\bf b}+{\bf v}_1) \\ C_1 ({\bf b}+{\bf v}_1)\end{pmatrix},
\begin{pmatrix} G_2 ({\bf b}+{\bf v}_2) \\ C_2 ({\bf b}+{\bf v}_2)\end{pmatrix},
\cdots,
\begin{pmatrix} G_s ({\bf b}+{\bf v}_s)\\ C_s ({\bf b}+{\bf v}_s)\end{pmatrix}\right]
 \end{align*} to be received at the decoder.
Given $C_i({\bf b}+{\bf v}_i)$ at the 
decoder, 
we can recover $Q_i({\bf b}+{\bf v}_i)$ from Remark \ref{rem:total_compression}. 

%
%
%
The decoder can then retrieve $({\bf v}_1,\cdots,{\bf v}_s)$ since
\begin{align*}
& Q_1({\bf b}+{\bf v}_1)+Q_2({\bf b}+{\bf v}_2)+ \cdots+Q_s({\bf b}+{\bf v}_s) \\
   = &Q_1({\bf v}_1)+\cdots+Q_s({\bf v}_s) & \mbox{(by (\ref{eqn:partition2}))} \\
   = &P\begin{pmatrix}
          {\bf v}_1      \\
          {\bf v}_2\\
\cdots       \\
          {\bf v}_s 
      \end{pmatrix} &   \mbox{(by (\ref{eqn:partition1}))}
\end{align*}
and $P$ is bijective 
over the set of all length-$sn$ vectors with weight $1$. 
 
      After knowing $({\bf v}_1,\cdots,{\bf v}_s)$, we can compute
   $G_1({\bf b}),\cdots,G_s({\bf b})$ and $C_1({\bf b}),\cdots,C_s({\bf b})$. 
This in turn gives us $T({\bf b})$ and $Y({\bf b})$, respectively, (the latter is again by Remark \ref{rem:total_compression}). 
So we have $R {\bf b}$. Since $R$ is invertible, we can get back ${\bf b}$ and thus all sources. 

The second claim is apparent by simple counting.
}
%
%
%
%
%
\end{IEEEproof} 

\begin{exam}[Generalized HCMS of three sources of length-$1$]

Let us revisit Remark \ref{rem:counter}.  
For the case $s=3$, $n=1$, and $M=3$.
consider
the Hamming matrix
$P= \begin{pmatrix}
     101\\
011
    \end{pmatrix}
=[ Q_1 Q_2 Q_3]
$
we must get $C_1=C_2=C_3=[1]$.
$\begin{pmatrix}
  C_1\\
C_2
 \end{pmatrix}=
\begin{pmatrix}
 1 \\
1
\end{pmatrix}
\Rightarrow Y=[1]$ and $T$ is  void and hence $Gi$ are void.
So $d_1=d_2=d_3=d=1$ and $d_1+d_2+d_3+(n-d)=M$. So from generalized HCMS, we get perfect compression with coding matrices 
$\begin{pmatrix} 1 \end{pmatrix} $, $\begin{pmatrix} 1 \end{pmatrix} $, 
and $\begin{pmatrix} 1 \end{pmatrix} $
just as in Remark \ref{rem:counter}.

\end{exam}

Note that even Generalized HCMS does not guaranties the existence of perfect compression as perfect compression may simply does not exist.
For example, there is no perfect compression for $s=3$, $n=5$, $M=3$ as shown in 
Proposition \ref{prop:a=2}.

\section{Universality of Generalized HCMS}
\label{sect:main}
 
We are to prove that every perfect compression for Hamming sources $S$ is equivalent to a generalized HCMS.
We say two perfect compressions are equivalent (denoted by $\sim$) to each other if and only if their null spaces can be converted to each other through the steps of the {\em null space shifting} as to be described in Lemma \ref{lem:allocate}. Since each step of null space shifting is invertible, the term ``equivalent'' is mathematically justify. The set of
perfect compression does form equivalent classes. The objective of this section is to show the following theorem.

\begin{thm}
\label{thm:main}
 Every perfect compression is equivalent to a generalized HCMS.
\end{thm}
 
To prove Theorem \ref{thm:main}, we will introduce and show several lemmas to achieve our goal. 

\begin{lem}
 Every 2-source perfect compression is equivalent to a Hamming code.
\label{lem:lem2}
\end{lem}


\begin{IEEEproof}
 Let $s=2$. If $(H_1, H_2)$ is a perfect compression, then we can let $N_1=\{{\bf 0}\}$ and $K=\mbox{null} (H_1)$ and form $(H'_1,H'_2)$ under Lemma 1. Having $\{{\bf 0}\}$ as null space, $H'_1$ can be any invertible $n \times n$ matrix and we can set $H'_1$ to the identity matrix without loss of generality. Meanwhile, $H'_2$ is a full rank $m \times n$ matrix with $2^m=n+1$.
Since the columns of $H_2$ must be nonzero and different from each other (because $i$-th column $=$ $j$-th column $\Leftrightarrow$
 $H'_2 {\bf e}_i = H'_2 {\bf e}_j$, 
{\color{black} where ${\bf e}_i = [\overset{i-1}{\overbrace{{\bf 0},\cdots,{\bf 0}}},1,\overset{n-i}{\overbrace{{\bf 0},\cdots,{\bf 0}}}]^T$ $\Leftrightarrow$ $(H'_1,H'_2)$ fails to compress $S$ }
because ${\bf e}_i$ and ${\bf e}_j$ inputted to encoder $2$ are no longer
distinguishable by the encoder's output),
$H'_2$ is unique up
to a permutation of columns. 
Therefore, 
$H'_2$ is a parity check matrix of the Hamming $(n,n-m)$ code.
Conversely, we can construct $(H_1,H_2)$ (up to their null spaces) from $(H'_1,H'_2)$ by Lemma \ref{lem:allocate}. That means any perfect compression is equivalent to Hamming code 
under Lemma \ref{lem:allocate}.
\end{IEEEproof}

\begin{lem}
\label{lem:lem3}
Given a perfect compression 
$(H_1', H_2', \cdots, H_s')$, there exists $(H_1, \cdots,H_s) \sim (H_1',\cdots,H_s')$ s.t.
$null H_1\cap...\cap null H_{i-1}\cap null H_{i+1}\cap...\cap null H_s={0}$ for $1 \le i<s$.
\end{lem}

Before we proceed with the proof of Lemma \ref{lem:lem3}, we will introduce a fact necessary for the proof as follows. 



\begin{fact}
For vector spaces $U,V,$ and $W$, it is easy to show that
\begin{equation}
(V+U)\cap W \subset (V\cap W)+U \mbox{ if } U\subset W. 
\label{eqn:3.7}
\end{equation}
\label{fact:subset} 
\end{fact}

\begin{IEEEproof}
Let $v\in V, u\in U$ that $v+u \in W$.
Then $u\in U\subset W \Rightarrow v\in W \Rightarrow v\in V\cap W$.
As a result $v+u \in (V\cap W) + U$.
\end{IEEEproof}

\ifthenelse{\commenton}
{
The first part of the proof is simple no matter how it may look. I never thought "null space 
shifting" can really be used for $s>3$.
}

\begin{IEEEproof}[Proof of Lemma \ref{lem:lem3}]
Let $R_i=\bigcap\limits_{1\le j \le s | j \neq i}\mbox{null} H'_j$ for $1 \le i<s$. We have
\begin{equation}
 R_i \subset null H_j', \mbox{ for } 1\le j\le s \mbox{ and } i\neq j,
\label{eqn:3.1}
\end{equation}
and
\begin{equation}
 R_i\cap R_k = \bigcap\limits_{1\le j\le s} null H_j'\overset{(a)}{=} {0} \mbox{ for } i \neq k,
\label{eqn:3.2}
\end{equation}
where (a) holds because the perfect
compression $(H_1',H_2',\cdots,H_s')$ must be injective and hence
the intersection of all of their null spaces must be {0}.

By \eqref{eqn:3.1} and \eqref{eqn:3.2}, there exist a space $N_s$  that we
can decompose
\begin{equation}
 null H_s'=N_s\oplus R_1\oplus...\oplus R_{s-1}
\label{eqn:3.3}
\end{equation}

Again by \eqref{eqn:3.1} and \eqref{eqn:3.2} together with 
Lemma \ref{lem:allocate},
we
can form an equivalent perfect compression by first moving the
whole $R_i$ from $null H_s'$ to $null H_i'$ for $i$ runs from $1$
to $s-1$ and get
\begin{equation}
   N_i=null H_i'\oplus R_i, 1\le i<s
\label{eqn:3.4}
\end{equation}
Then if we let $H_j$ be a surjective matrix with
\begin{equation}
 null H_j=N_j \mbox{ for } 1\le j\le s,
\label{eqn:3.5}
\end{equation}
we have $(H_1,\cdots,H_s)\sim(H_1',\cdots ,H_s')$.
 
Now let $L_i=\bigcap_{1\le j\le s| j\neq i}N_j$ for $1\le i<s$,
we still need to show $L_i={0}$.

By symmetry, all we need to show
is $L_1=N_2\cap N_3\cap\cdots \cap N_{s-1}\cap N_s=0$. 
By \eqref{eqn:3.4}, we have $N_2\subset null H'_2 \oplus R_2$. 
Suppose
$N_2\cap \cdots \cap N_k \subset (null H_2'\cap \cdots \cap null H'_k) + (R_2+ \cdots +R_k)$
for a $k<s-1$. Then
$N_2\cap \cdots \cap N_{k+1}$ is a subset of
\begin{equation}
\begin{split}
 ((nullH_2'\cap \cdots \cap null H'_k) + (R_2+ \cdots +R_k)) \\\cap (null H'_{k+1}+R_{k+1})
\end{split}
\label{eqn:3.6}
\end{equation}
by induction hypothesis.

By \eqref{eqn:3.1} $R_2+\cdots+R_k\subset null H'_{k+1}\subset null H'_{k+1}+R_{k+1}$. So
we can apply Fact \ref{fact:subset} 
on \eqref{eqn:3.6} and thus obtain
$N_2 \cap\cdots\cap N_{k+1}$ as a subset of
$((nullH_2'\cap..\cap null H'_k) \cap (null H'_{k+1}+R_{k+1})+ (R_2+\cdots+R_k))$.
Apply Fact \ref{fact:subset} 
once more with 
$V=null H'_{k+1}, U=R_{k+1}, W=null H_2' \cap \cdots \cap null H'_k$,
we get (c.f. \eqref{eqn:3.1} for $U\subset W$)
$N_2\cap\cdots\cap N_{k+1}$ is a subset of
$(null H_2' \cap \cdots \cap null H'_{k+1})+R_2+R_3+\cdots+R_{k+1}$.
By induction we get
$N_2\cap \cdots \cap N_{s-1}\subset (null H_2'\cap \cdots \cap null H'_{s-1})+R_2+\cdots+R_{s-1}$.
                              
Lastly, $N_2\cap \cdots \cap N_s$
\begin{align*}
\overset{(a)}{\subset} & ((null H_2'\cap \cdots \cap null H'_{s-1})+R_2+\cdots+R_{s-1} ) \\ 
& \qquad \cap 
null H'_{s} \\
\overset{(b)}{ \subset} & (null H'_2\cap\cdots\cap null H'_s) + R_2 +\cdots+R_{s-1} \\  
 \overset{(c)}{=} & R_1+R_2+\cdots+R_{s-1}, 
\end{align*}
where (a) is due to  $N_s\in null H'_{s}$ (c.f. \eqref{eqn:3.3}), (b) is due to 
Fact \ref{fact:subset}, and (c) is from the definition of $R_1$.
 
Thus, 
$ N_2\cap \cdots \cap N_s 
\subset  (R_1+\cdots+R_{s-1}) \cap N_s  
                                =  \{0\}, 
$
where the last equality is from the construction of $N_s$ (c.f. \eqref{eqn:3.3}).
\end{IEEEproof}


\begin{lem}
\label{lem:lem4}
 Given the coding matrices, $(H_1,\cdots,H_s)$,  of a perfect $(s,n,M)$-compression, 
we can form a perfect $(2,sn,M+(s-1)n)$-compression with coding matrices $(X,J)$, where
\begin{equation}
\small
X=
\begin{pmatrix}
 I & I & 0 & 0 &\cdots & 0& 0& 0 \\
     0& I & I & 0 &\cdots & 0&0&0 \\
    &&& \cdots \\
      0& 0& 0& 0 & \cdots & 0& I& I
\end{pmatrix}
\label{eqn:4.1}
\end{equation}
is 
a $(s-1)n \times sn$ matrix, and
\begin{equation}
\small
J= 
\begin{pmatrix}
H_1 & 0 & \cdots & 0\\
      0 & H_2 & \cdots & 0\\
& & \cdots \\
      0 & 0 & \cdots & H_s
\end{pmatrix}
\label{eqn:4.2}
\end{equation}
is a $M \times sn$ matrix, 
and $I$ denotes the $n\times n$ identity matrix.
 \end{lem}

 
{\color{black}
 
\begin{IEEEproof}
 Since $2^n(sn+1)=2^M$ implies $2^{sn}(sn+1)=2^{M+(s-1)n}$, we only need to show how to
 retrieve the input vectors. Let us decompose any pair of the input vectors for $X$ and $J$, respectively, into
$
\begin{pmatrix}
{\bf b}_1 \\
{\bf b}_2 \\
\cdots \\
{\bf b}_s
\end{pmatrix}
$
and
 $
\begin{pmatrix}
{\bf b}_1+{\bf v}_1 \\
\cdots \\
{\bf b}_s+{\bf v}_s 
\end{pmatrix},
$ 
 where ${\bf b}_i$'s are $n$-entry vectors, ${\bf v}_i$'s are also $n$-entry vectors but restricted to
the condition $|{\bf v}_1|+\cdots+|{\bf v}_s| \le 1$, where 
$|\cdot|$ maps an element in ${\mathbb Z}_2^n$ to its norm in $\mathbb Z$ by counting the number of nonzero components.

 
From the output of $X$, we will get
${\bf b}_1+{\bf b}_2, {\bf b}_2+{\bf b}_3,\cdots,{\bf b}_{s-1}+{\bf b}_s$. 
Thus we can obtain 
${\bf b}_1+{\bf b}_2, {\bf b}_1+{\bf b}_3,\cdots,{\bf b}_1+{\bf b}_s$ and 
$H_2({\bf b}_1+{\bf b}_2), H_3({\bf b}_1+{\bf b}_3),\cdots,H_s({\bf b}_1+{\bf b}_s)$. 

From the output of $J$, we will obtain
$H_2({\bf b}_2+{\bf v}_2),H_3({\bf b}_3+{\bf v}_3),\cdots,H_s({\bf b}_s+{\bf v}_s)$ and $H_1({\bf b}_1+{\bf v}_1)$.
Combining the results, we get
$H_1({\bf b}_1+{\bf v}_1), H_2({\bf b}_1+{\bf v}_2),\cdots,H_s({\bf b}_1+{\bf v}_s)$.
Since $(H_1,\cdots,H_s)$ is a perfect compression, we can compute
${\bf b}_1, {\bf v}_1, {\bf v}_2,\cdots,{\bf v}_s$. Together with the output of $X$, we can retrieve all
${\bf b}_1,{\bf b}_2,\cdots,{\bf b}_s$ and ${\bf v}_1,{\bf v}_2,\cdots,{\bf v}_s$.
\end{IEEEproof}
}


Before we finally proceed to the proof of Theorem \ref{thm:main}, we need to present one more fact. 

\begin{fact}
For vector spaces $V$, $U$, and $W$,
$(V \oplus U) \cap W= (V\cap W)\oplus U$ if $U\subset W$.
 \label{fact:directsum}
\end{fact}

{\color{black}
\begin{IEEEproof}
From \eqref{eqn:3.7}, we know that $(V+U)\cap W \subset (V\cap W)+U$. 
Now let ${\bf v}\in V\cap W$, ${\bf u}\in U\subset W$.
Then ${\bf v}+{\bf u} \in W$ and ${\bf v}+{\bf u} \in V+U$. Therefore
$(V\cap W)+U \subset (V+U)\cap W$.
Lastly, we notice that $V\cap U=(V\cap W)\cap U$ as $U\subset W$.
Hence $V\cap U=\{0\}$ iff $(V\cap W)\cap U=\{0\}$, that justifies the direct sum signs.
\end{IEEEproof}
}

 
\begin{IEEEproof}[Proof of Theorem \ref{thm:main}]

By Lemma \ref{lem:lem3},
we can restrict our attention only to 
perfect 
compression 
whose coding matrices $(H_1,..,H_s)$ satisfy
\begin{equation}
null H_1\cap null H_2\cdots\cap null H_{i-1} \cap null H_{i+1}\cap \cdots\cap null H_s={0}   
\label{eqn:5.1}
\end{equation}
for $i \neq s$ 
without loss of generality. 

We can also generate $X$ and $J$ according to \eqref{eqn:4.1} and \eqref{eqn:4.2}. Then Lemma \ref{lem:lem4} shows that $(X,J)$ is a
perfect $(2,sn,M+(s-1)n)$-compression.
Therefore $null X \cap null J={0}$. Then Lemma \ref{lem:allocate} 
tells us that two surjective matrices with null spaces $\{0\}$ and $null X\oplus null J$, respectively, are also a perfect compression. By the proof of Lemma \ref{lem:lem2}, the first matrix is an invertible matrix and the second one is a $(M+(s-1)n-sn=)M-n$-bit Hamming matrix $P$.
 
We have
\begin{equation}
null P=null X \oplus null J                                                             
\label{eqn:5.2}
\end{equation}
and
{\color{black}
\begin{equation}
\begin{split}
null X=\{({\bf c},\cdots,{\bf c})^t | {\bf c}\in Z_2^n\}; \\ null J=\{({\bf n}_1,\cdots,{\bf n}_s)^t | {\bf n}_i\in null H_i\}.
\end{split}
\label{eqn:5.3} 
\end{equation}
}
Partition $P$ into
\begin{equation}
 P=[Q_1 Q_2\cdots.Q_s],
\label{eqn:5.4}
\end{equation}
such that $Q_i$ is a $(M-n) \times n$ matrix. We have
\begin{equation}
Q_1+Q_2+\cdots+Q_s=0
\label{eqn:5.5}
\end{equation}
because $null X \subset null P$.

Secondly, $ null J\subset null P$ implies
\begin{equation}
null H_j \subset null Q_j
 \label{eqn:5.6}
\end{equation}
for $1 \le j\le s$. 
Moreover, let
\begin{equation}
 b_j\in null Q_j,
\label{eqn:5.7}                                                                        \end{equation}
{\color{black}
we have $({\bf 0},\cdots,{\bf 0},{\bf b}_j,{\bf 0},\cdots,{\bf 0})^t \in null P$ and we can decompose it into
   $({\bf c},\cdots,{\bf c}) + ({\bf c},\cdots,{\bf c},{\bf c}+{\bf b}_j,{\bf c},\cdots,{\bf c})$ by \eqref{eqn:5.2} with
\begin{equation}
 ({\bf c},\cdots,{\bf c},{\bf c}+{\bf b}_j,{\bf c},\cdots,{\bf c})\in null J
\label{eqn:5.8}
\end{equation}
So ${\bf c}\in null H_1\cap null H_2\cap \cdots \cap null H_{j-1} \cap null H_{j+1}\cap \cdots \cap null H_s$
and ${\bf c}+{\bf b}_j\in null H_j$. By \eqref{eqn:5.1}, we have ${\bf c}={\bf 0}$ if $j\neq s$ and ${\bf b}_j\in null H_j$. Hence $null Q_j\subset null H_j$. Together with \eqref{eqn:5.6}, we get }
\begin{equation}
null H_j= null Q_j \mbox{ for } j\neq s.
\label{eqn:5.9} 
\end{equation}

Recall that $Y$ 
is a row basis matrix 
of
\begin{equation}
\left[ 
\begin{matrix}
Q_1 \\
 Q_2 \\
  ... \\
 Q_{s-1} 
\end{matrix}
\right]                                                                        
\label{eqn:5.12} 
\end{equation}
and hence
\begin{equation}
null Y= null Q_1\cap...\cap null Q_{s-1}=null H_1\cap ...\cap null H_{s-1}
\label{eqn:5.13}
\end{equation}
by \eqref{eqn:5.9}. Now we will show that $null Q_s = null H_s \oplus null Y$.


{\color{black}
{\em $null Q_s  \subset null H_s \oplus null Y$}:
Equations \eqref{eqn:5.7} and \eqref{eqn:5.8} are true for all $j$. 
In particular, when $j=s$ and let ${\bf b}_s\in null Q_s$, 
we have 
${\bf b}_s={\bf c}+{\bf b}_s+{\bf c}$, where
${\bf c}\in null H_1\cap \cdots \cap null H_{s-1}=null Y$ 
and
${\bf c}+{\bf b}_s\in null H_s$ by \eqref{eqn:5.8}.
Therefore, ${\bf b}_s\in null H_s+ null Y$.
Since $null H_s \cap null Y=null H_1\cap ...\cap null H_s={{\bf 0}}$,
we have $null Q_s\subset null H_s\oplus null Y$.
 

{\em $null H_s \oplus  null Y \subset null Q_s$}: 
Given ${\bf b}_s\in null H_s$ and ${\bf c}\in null Y$,
$({\bf 0},{\bf 0},\cdots,{\bf 0}, {\bf b}_s+{\bf c})^t=({\bf c},\cdots,{\bf c})^t+({\bf c},\cdots,{\bf c},{\bf b}_s)^t \in null P$ (c.f. \eqref{eqn:5.2}, \eqref{eqn:5.3})
because $({\bf c},\cdots,{\bf c})\in null X$ and $({\bf c},\cdots,{\bf c},{\bf b}_s)$ in $null J$.
Thus ${\bf 0}=P({\bf 0},\cdots,{\bf 0},b_s+{\bf c})^t=Q_s({\bf b}_s+{\bf c})$. Hence 
${\bf b}_s + {\bf c} \in null Q_s$ and thus we have
\begin{equation}
null Q_s=null H_s\oplus null Y
\label{eqn:5.14}                                                       
\end{equation}
}


Let $A$ be a subspace of $Z_2^n$ such that
\begin{equation}
null H_s\oplus null Y\oplus A=Z_2^n.                                     
 \label{eqn:5.15}
\end{equation}

Let $T$ be a surjective matrix with
\begin{equation}
null T =null H_s\oplus A.
\label{eqn:5.16}                                                              
\end{equation}

\ifthenelse{\commenton}
{
($T$ is the $T$ in Generalized HCMS, so $T$, $Y$ and $C_i$ 
all have shown up and the story is coming to the end)
($T$ is a void matrix in case $null T=0$.)
}


We have
\begin{equation}
null T\oplus null Y=Z_2^n              
 \label{eqn:5.17}
\end{equation}

Being a row basis matrix matrix (c.f. \eqref{eqn:5.12}), $Y$ is also surjective.
We have
\begin{equation}
R=\left[ \begin{matrix}
           Y \\T
          \end{matrix}
 \right]
\label{eqn:5.18} 
\end{equation}
 is an $n \times n$ invertible matrix.                             

\ifthenelse{\commenton}{
(\# of row of $Y = n - dim (null Y) = dim (null T) \Rightarrow R $ is $n\times n$.
 $null R=null Y \cap null T={0} \Rightarrow R$ is injective and square $\Rightarrow$ bijective )
 }

{\color{black}
$null Q_s \cap null T=(null H_s\oplus null Y)\cap null T$. 
Since $null H_s\subset null T$   (c.f. \eqref{eqn:5.16}), we can apply 
Fact \ref{fact:directsum}
to obtain  
\begin{align}
 null Q_s \cap null T &=(null T \cap null Y)\oplus null H_s \\ \nonumber  
                           &={0}\oplus null H_s \mbox{ (c.f. \eqref{eqn:5.17}) }  \\ \nonumber  
                           &= null H_s
\end{align}
}

Denote $C_j$ as a row basis matrix of $Q_j$ for all $j$,                                   
\ifthenelse{\commenton}
{
(I don't like the name of row basis matrix. It is too close to perfect compression.)
}
then we have
\begin{equation}
null C_j=null Q_j 
 \label{eqn:5.11}
\end{equation}
for all $j$.       

{\color{black}
Since $Q_s=Q_1+Q_2+...+Q_{s-1}$ (c.f. \eqref{eqn:5.5}) and $C_s$ is a row basis matrix
of $Q_s$ (c.f. \eqref{eqn:5.11}), we have
\begin{equation}
row C_s=row Q_s \subset  row Y
 \label{eqn:5.20} 
\end{equation}}
(c.f. \eqref{eqn:5.12}).                       
Then
$\left[
\begin{matrix}
T \\ C_s 
\end{matrix}
\right]$ is a surjective matrix because its row vectors are linear independent, thanks to
\eqref{eqn:5.18} and \eqref{eqn:5.20}.
 
Therefore, $\left(C_1, C_2,\cdots, C_{s-1}, \left[\begin{matrix} T \\C_s \end{matrix}\right] \right)$ is a perfect compression as its
encoding matrices are all surjective and have the same null spaces of
the perfect compression $(H_1,\cdots,H_s)$'s (c.f. Lemma \ref{lem:allocate}). 
Moreover
$\left(C_1, C_2,\cdots, C_{s-1}, \left[\begin{matrix} T \\C_s \end{matrix}\right] \right)$ is a generalized HCMS which is equivalent to $(H_1,\cdots,H_s)$.
 
\end{IEEEproof}

\section*{Appendix}
\begin{IEEEproof}[Proof of Proposition \ref{prop:a=2}]

Denote $S$ as the set containing all $3$-terminal Hamming source of length $5$. 
Let $H_1,H_2,H_3$ be the three parity check matrices. 
Our strategy is to limit the null spaces of them by the fact that 
all elements in $S$ need to have distinct syndromes.
The limitation will eventually kill the possibility of the existence of
$H_1,H_2,H_3$.

Denote the null set of a matrix $A$ as $\mbox{null}(A)=\{{\bf u} | A {\bf u}=
{\bf 0}\}$, where $\bf 0$ is an all zero vector. Further, denote ${\bf e}_i$ as
the length-$5$ binary column vector that has $i^{th}$ component equal to $1$ and
the rest of its components equal to zero. 

We may assume the number of row in $H_1$ is smaller than or equal to the other's. In other 
words, $H_1$ has at most $3$ rows.
Hence $\mbox{null} (H_1)$ has at least two
degrees of freedom. 
Regardless the values of $H_2$ and $H_3$,
 $\mbox{null} (H_1)$ cannot contain any ${\bf e}_i$. Otherwise, 
both 
$({\bf e}_i,{\bf
0},{\bf 0})$ and $({\bf 0},{\bf 0},{\bf 0})$ that are in $S$  will get the same outputs 
$({\bf y}_1,{\bf y}_2,{\bf y}_3)=({\bf
0},{\bf 0},{\bf 0})$.
Similarly,
$\mbox{null} (H_1)$ cannot contain ${\bf e}_i+{\bf e}_j$ neither, otherwise both
$({\bf e}_i,{\bf 0},{\bf 0})$ and $({\bf e}_j,{\bf 0},{\bf 0})$ (in $S$) will get the
same
outputs because $H_1 {\bf e}_i=H_1 {\bf e}_j$. 
Thus $\mbox{null} (H_1)$ can only be $\mbox{span}({\bf e}_i+{\bf e}_j+{\bf e}_k,
{\bf e}_i+{\bf e}_m+{\bf e}_n)$, where the letters $i,j,k,m,n$ are different to
each others.
     i.e. $\mbox{null} (H_1)=\{{\bf 0}, {\bf e}_i+{\bf e}_j+{\bf e}_k, {\bf
e}_i+{\bf e}_m+{\bf e}_n, {\bf e}_j+{\bf e}_k+{\bf e}_m+{\bf e}_n\}$ for some
$i,j,k,m,n$. Other structures such as higher dimension will contain forbidden elements. 
As the dimension of the null spaces of 
 $ 1 \times 5$ and $2 \times 5$ matrices are all
greater than $2$, $H_1$ has to have at least $3$ rows. 
Thus both $H_2$ and $H_3$ also have three rows. It 
excludes the possibility of perfect asymmetric SW codes (at rate [$2/5,3/5,4/5$], for example). 
So we can focus only on the symmetric case from now on. 

From the above discussion, $\mbox{null} (H_1)$ has to contain 
 ${\bf 0}$, two ``3{\bf e}'' vectors and one ``4{\bf e}'' vector, and no
more.
Similarly, $\mbox{null} (H_2)$ and $\mbox{null} (H_3)$ get the same structure.
 
 
Without lose of generality, we can write
$\mbox{null} (H_1)=\{{\bf 0}, {\bf e}_1+{\bf e}_2+{\bf e}_3, {\bf e}_1+{\bf
e}_4+{\bf e}_5, {\bf e}_2+{\bf e}_3+{\bf e}_4+{\bf e}_5\}$. 
Suppose $\mbox{null} (H_2)$ contain ${\bf e}_1+{\bf e}_2+{\bf e}_3$, then of
course $\mbox{null} (H_3)$ cannot
contain ${\bf e}_1+{\bf e}_2+{\bf e}_3$.
Otherwise, $({\bf 0},{\bf 0},{\bf 0}) \in S$ and $({\bf e}_1+{\bf e}_2+{\bf e}_3,
{\bf e}_1+{\bf e}_2+{\bf e}_3, {\bf e}_1+{\bf e}_2+{\bf e}_3) \in S$ get the
same output $({\bf 0},{\bf 0},{\bf 0})$.
But $\mbox{null} (H_3)$ cannot contain ${\bf e}_i+{\bf e}_j+{\bf e}_k$, $i,j \in
\{1,2,3\}$; $k \in \{4,5\}$ $(i \neq j)$. Otherwise
$({\bf e}_1+{\bf e}_2+{\bf e}_3, {\bf e}_1+{\bf e}_2+{\bf e}_3, {\bf e}_i+{\bf
e}_j) \in S$ and $({\bf 0}, {\bf 0}, {\bf e}_k) \in S$ share the same output as
well. 
So the ``3{\bf e}'' vectors of $\mbox{null} (H_3)$ can only be two of ${\bf e}_1+{\bf e}_4+{\bf
e}_5$, ${\bf e}_2+ {\bf e}_4 +{\bf e}_5$, and ${\bf e}_3+{\bf e}_4+{\bf e}_5$.
Unfortunately, any pair of them sum up to
a ``2{\bf e}'' vector 
instead of a ``4{\bf e}''
vector. 
Therefore,
there cannot 
be a common ``3{\bf e}'' vector shared between any pair of
the null spaces of $H_1, H_2$, and $H_3$.


Now, 
suppose $\mbox{null} (H_2)$ contains the same ``4{\bf e}'' vector ${\bf
e}_2+{\bf e}_3+{\bf e}_4+{\bf e}_5$ as $\mbox{null} (H_1)$ does. Then
$\mbox{null} (H_3)$ cannot contain any
``4{\bf e}'' vector.
Let the ``4{\bf e}'' vector of $\mbox{null} (H_3)$ be ${\bf e}_1+{\bf e}_2+{\bf
e}_3+{\bf e}_4+{\bf e}_5-{\bf e}_j$, $j\in \{2,3,4,5\}$.
Then $({\bf e}_2+{\bf e}_3+{\bf e}_4+{\bf e}_5, {\bf e}_2+{\bf e}_3+{\bf
e}_4+{\bf e}_5, [1,1,1,1,1]^T)$ and
        $({\bf 0}                  , {\bf 0}                   , {\bf e}_j)$
shares the same syndrome.
Thus, any pair of null spaces of $H_1, H_2$, and $H_3$ cannot share a common
``4{\bf e}'' vector as well.
 
Hence, without loss of generality, we can write
          $\mbox{null} (H_2)=\{{\bf 0}, {\bf e}_2+{\bf e}_1+{\bf e}_4, {\bf
e}_2+{\bf e}_3+{\bf e}_5, {\bf e}_1+{\bf e}_3+{\bf e}_4+{\bf e}_5\}$.
Then, there are only three different possibilities for the ``4{\bf e}'' vector of $\mbox{null}(H_3)$:
\begin{enumerate}
\item
$ {\bf e}_1+{\bf e}_2+{\bf e}_4+{\bf e}_5$;
\item
$ {\bf e}_1+{\bf e}_2+{\bf e}_3+{\bf e}_5$;
\item
$ {\bf e}_1+{\bf e}_2+{\bf e}_3+{\bf e}_4$.
\end{enumerate}


Case 1 does not work because
          $ ({\bf e}_1+{\bf e}_4+{\bf e}_5, {\bf e}_1+{\bf e}_3+{\bf e}_4+{\bf
e}_5, {\bf e}_1+{\bf e}_4+{\bf e}_5) \in S$
and     $ ({\bf 0}             , {\bf 0}                  ,{\bf e}_2           
) \in S$
shares the same syndrome.
 
Case 2 does not work neither because
          $({\bf e}_1+{\bf e}_2+{\bf e}_3, {\bf e}_1+{\bf e}_2+{\bf e}_3+{\bf
e}_5, {\bf e}_1+{\bf e}_2+{\bf e}_3+{\bf e}_5) \in S$
and   $ ( {\bf 0}             ,  {\bf e}_1                ,{\bf 0}              
   ) \in S$
share the same syndrome.

Finally, case 3 fails as well since 
          $({\bf e}_1+{\bf e}_2+{\bf e}_3, {\bf e}_1+{\bf e}_2+{\bf e}_3+{\bf
e}_4, {\bf e}_1+{\bf e}_2+{\bf e}_3+{\bf e}_4) \in S$
and   $ ( {\bf 0}             ,  {\bf e}_3                ,{\bf 0}              
   ) \in S$
share the same syndrome.
\end{IEEEproof}

\bibliographystyle{IEEEtran}
\bibliography{../../ref}

\end{document}